\documentclass{article}

\usepackage{PRIMEarxiv}

\usepackage[utf8]{inputenc} 
\usepackage[T1]{fontenc}    
\usepackage{nicefrac}       
\usepackage{microtype}      
\usepackage{lipsum}
\usepackage{fancyhdr}       

\usepackage{amsmath,amssymb,amsfonts}
\usepackage{graphicx}
\usepackage{textcomp}
\usepackage{xcolor}
\usepackage{tipa}
\usepackage{rotating}
\usepackage{booktabs}

\usepackage{subcaption}
\usepackage{lineno}
\usepackage{soul}
\usepackage{multirow}
\usepackage{hyperref}

\usepackage{tablefootnote}

\usepackage{xurl}

\title{Real-time Detection of AI-Generated Speech for DeepFake Voice Conversion}

\author{
  Jordan J. Bird, Ahmad Lotfi \\
  Nottingham Trent University \\
  Nottingham, UK\\
  \texttt{\{jordan.bird, ahmad.lotfi\}@ntu.ac.uk} \\
  }

\begin{document}
\maketitle

\begin{abstract}
There are growing implications surrounding generative AI in the speech domain that enable voice cloning and real-time voice conversion from one individual to another. This technology poses a significant ethical threat and could lead to breaches of privacy and misrepresentation, thus there is an urgent need for real-time detection of AI-generated speech for DeepFake Voice Conversion. To address the above emerging issues, the DEEP-VOICE dataset is generated in this study, comprised of real human speech from eight well-known figures and their speech converted to one another using Retrieval-based Voice Conversion. Presenting as a binary classification problem of whether the speech is real or AI-generated, statistical analysis of temporal audio features through t-testing reveals that there are significantly different distributions. Hyperparameter optimisation is implemented for machine learning models to identify the source of speech. Following the training of 208 individual machine learning models over 10-fold cross validation, it is found that the Extreme Gradient Boosting model can achieve an average classification accuracy of 99.3\% and can classify speech in real-time, at around 0.004 milliseconds given one second of speech. All data generated for this study is released publicly for future research on AI speech detection.
\end{abstract}

\keywords{DeepFake Detection \and Generative AI \and Speech Recognition \and Audio Signal Processing \and Voice Cloning}

\section{Introduction}
The implications of generative Artificial Intelligence (AI) in recent years are rapidly growing in importance. State-of-the-art systems capable of converting a speaker's voice to another in real-time via a microphone and sophisticated deep learning models. The ability to clone an individual's speech and use it during an online or phone call is no longer science fiction, and is possible using consumer-level computing technology. 

Although this technology may prove attractive for entertainment purposes, advancements in the field pose a significant security threat. Human beings use voice as a method of recognising others in social situations and often go unquestioned. Voice recognition is also used for biometric authentication, and thus voice conversion could be used unethically to breach privacy and security. In this case, the potential for misrepresentation and identity theft are enabled, which requires immediate solutions from the scientific literature. 

The scientific contributions of this work are threefold: first, the provision of an original audio classification dataset comprised of 8 well-known public figures, with real audio collected from the internet and AI-generated speech via Retrieval-based Voice Conversion (RVC). Second, the statistical analysis of extracted audio features to explore which sets of features are statistically significant given the classification of human or AI-generated speech. Third, the hyperparameter optimisation of statistical Machine Learning (ML) models towards improving accuracy and inference time, in order to achieve real-time recognition of AI-generated speech. The real-time models presented by this study are important for real-world use, and could be used, for example, to provide a warning system for individuals on phonecalls or in conference calls, where a synthetic voice may be part of the conversation with nefarious aims. 

This research article also contributes the DEEP-VOICE\footnote{The data used in this study is available from:\\ \url{https://www.kaggle.com/datasets/birdy654/deep-voice-deepfake-voice-recognition}} dataset to enable the analysis of AI-generated speech. The datasets collected and generated for this study are released publicly for the research community to facilitate interdisciplinary work on the analysis and recognition of AI-generated speech patterns. This study also explores ML-based countermeasures against synthetic speech impersonation.

The remainder of this study is as follows, Section \ref{sec:background} first explores the scientific literature relevant to this study before Section \ref{sec:method} outlines the methodology followed by the experiments in this work. The experimental results are then presented and discussed in Section \ref{sec:results}. Finally, this work is concluded along with suggestions for future work due to the findings of these studies in Section \ref{sec:conclusion}. 

\section{Background}
\label{sec:background}
This section explores related work in the field, providing background information on deepfakes and synthetic media, as well as their detection. Finally, this section discusses how the literature has influenced the methodological design of the experiments in this work. 

Deep Learning Fakes (DeepFakes) describe a category of algorithms which can generate synthetic media with the purpose of replacing an individual's likeness with another \cite{juefei2022countering}. This form of synthetic media leads to many social, ethical, and legal issues regarding trustworthiness in data, and can be used to portray a human being doing or saying something that they did not, in reality \cite{banks2018deepfakes,waldrop2020synthetic}. The most common examples of DeepFakes are through images, oftentimes replacing one's face with another where the second person is engaging in an activity that the victim did not \cite{borel2018clicks}. Truby and Brown describe digital human clones as the \textit{Holy Grail of Artificial Intelligence} \cite{truby2021human}, noting that human behaviour becoming a commodity has led to the evolution of digital clones being sought on an industrial level. Truby and Brown's legal and ethical study also noted that Europe's General Data Protection Regulation provides a legislative example for other juristictions to prevent unauthorised digital cloning of individuals given their harmful potential. Rapid advances in synthetic media have enabled many other forms of digital cloning, and this study focusses on the replication of a voice and its detection. 

Audio visual cloning was notably suggested in 2001 by Beard \cite{beard2001clones} as a method to replicate an actor within a film after they have died. This could be in the form of an insurance policy given that an individual may not finish a film, or to license the likeness of an actor within their estate. Thus, this concept is closely related to the idea of \textit{digital immortality}, which is the digital replication of a posthumous human being \cite{meese2015posthumous}. 

In 2018, Buzzfeed used an application called FakeApp to replace the face and voice of comedian Jordan Peele with that of US President Barack Obama \cite{agarwal2019protecting}. Although presenting as a comedy sketch, the video brought much attention to the realistic nature of synthetic media. It is noted that only several minutes of example speech were needed to transfer style. As of the time of writing, five years of research have continued the field, with enhanced methods passing into and beyond the uncanny valley \cite{wells2022s,coburn2022enhanced,dale2022voice}.

Google introduced a speech synthesis model named Tacotron in 2017 \cite{wang2017tacotron}, and subsequent research introduced prosody \cite{skerry2018towards} (intonation and stress), style control\cite{wang2018style}, multi-speaker synthesis \cite{jia2018transfer}, among many more advances. Noting this growing increase in quality and, as such, misplaced trust, researchers also explore how synthesised speech can be detected. In 2022, Lim, Suk-Young and Lee \cite{lim2022detecting} proposed Convolutional and Temporal models for detection. This study showed that the speech generated from Tacotron sometimes showed a relatively flatter spectrogram without randomness as could be observed in human speech. Researchers conjecture that this may be due to the absence of accents in LJSpeech (the base tacotron dataset) rather than an inability of the pipeline itself. The study showed that Convolutional Neural Networks (CNNs) and Long-Short-Term-Memory (LSTM) neural networks could score around 97-99\% accuracy when dealing with recognition. It must be noted that temporal convolutional approaches are relatively computationally expensive. 

Similarly, studies in \cite{chen2020generalization} showed that a residual CNN achieved the lowest error rate for the ASVspoof 2019 challenge at 4.04\%, and later this was reduced to 1.26\%. Mcuba et al. \cite{mcuba2023effect} also explored a similar problem, and implemented various convolutional neural networks to learn from images generated from chromagrams, spectrograms, mel-spectrum, and mel-frequency cepstral coefficients. This study found that a VGG-16 CNN with an Adadelta optimiser was around 85.91\% accurate for deepfake detection. Furthermore, prior results found several models to score in the 40-60\% range, suggesting that the recent advancements in deepfake speech lead to difficulty in their detection. Conti et al. \cite{conti2022deepfake} propose that high-level features from Speech Emotion Recognition (SER) are indicative of generative speech. The experiments were tested on the ASVSpoof2019 dataset \cite{wang2020asvspoof} and showed that transfer learning from the SER model enabled better classificaiton results for neural speech detection. 

Inspired by the literature, this study proposes the use of chromagrams, spectrograms, mel-spectrum and mel-frequency cepstral coefficients similarly to \cite{mcuba2023effect}, while also considering the problems of computational complexity. Algorithms such as CNN and LSTM, and their fused counterpart, show effectiveness but are complex. Thus, for the consumer, these algorithms will likely not infer data in real-time. For that reason, this study will also focus on the optimisation of statistical algorithms, considering the inference time alongside ability.

\section{Method}
\label{sec:method}
This section provides an overview of the methodology followed in this work. An overview of data collection and preprocessing is provided before details on the DeepFake voice conversion approach employed. Following this, details of the ML models and their optimisation are provided. Finally, details of the hardware and software used for these experiments are described for the purposes of replicability. 

\begin{figure}[t]
    \centering
    \includegraphics{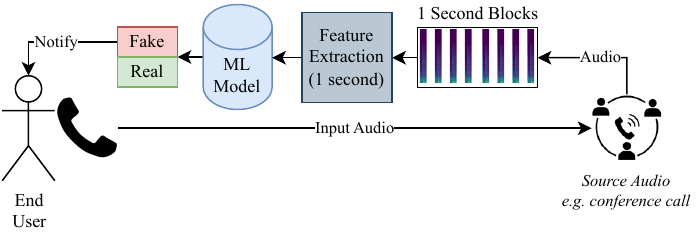}
    \caption{Usage of the real-time system. The end user is notified when the machine learning model has processed the speech audio (e.g. a phone or conference call) and predicted that audio chunks contain AI-generated speech.}
    \label{fig:usage-diagram}
\end{figure}

The research question for this study is how to detect Synthetic Speech in real-time and inform the end-user accordingly. The diagram illustrated in Figure \ref{fig:usage-diagram} shows a use case that the proposed system could be implemented. The source audio, such as a phone or conference call, is processed and classified. If the audio is predicted to contain AI-generated speech, the end user is notified. 

\subsection{Data Collection and Preprocessing}
Eight individuals are selected, each with a source for the basis of real speech and data to be converted to AI-generated speech. Table \ref{tab:dataset-length} provides the speech sources. In total, 62 minutes and 22 seconds of speech are collected from eight individuals. Audio tracks are limited to a maximum of ten minutes. As can be observed, some of the tracks have more background noise than others, such as cheering from supporters during Presidential victory speeches. Some of the speech tracks are of production-level quality, such as the \textit{Stepping Down Monologue} spoken by Linus Sebastian, whereas others are of lower quality, an example being Elon Musk's \textit{Commencement Speech} which was recorded at a distance without studio-quality hardware. The tracks are chosen based on these attributes to provide variation within the dataset. 

\begin{table}[t]
\caption{Data collected for training, validation, and unseen testing for the experiments in this work (sorted alphabetically by surname). Audio segments are cropped to a maximum of ten minutes. }
\label{tab:dataset-length}
\centering
\footnotesize
\begin{tabular}{llr}
\hline
\textbf{Individual}                                        & \textbf{Source}         & \textbf{Length (MM:SS)} \\ \hline
Joe Biden   & Victory Speech\tablefootnote{Victory Speech by Joe Biden:  \url{https://www.youtube.com/watch?v=1AfNYztas2c} Last accessed: 07/23}          & 10:00                   \\
Ryan Gosling  & Golden Globes Speech\tablefootnote{Golden Globes Speech by Ryan Gosling: \url{https://www.youtube.com/watch?v=K8JLyUW\_MSw} Last accessed: 07/23}    & 1:33                    \\
Elon Musk    & Commencement Speech\tablefootnote{Commencement Speech by Elon Musk: \url{https://www.youtube.com/watch?v=MxZpaJK74Y4} Last accessed: 07/23}     & 10:00                   \\
Barack Obama  & Victory Speech\tablefootnote{Victory Speech by Barack Obama: \url{https://www.youtube.com/watch?v=IeCY-jKpoZ0} Last accessed: 07/23}          & 10:00                   \\
Margot Robbie  & BAFTAs Speech\tablefootnote{BAFTAs Speech by Margot Robbie:  \url{https://www.youtube.com/watch?v=-JA3\_QBfjG8} Last accessed: 07/23}           & 1:19                    \\
Linus Sebastian    & Stepping Down Monologue\tablefootnote{Stepping Down Monologue by Linus Sebastian: \url{https://www.youtube.com/watch?v=0vuzqunync8} Last accessed: 07/23} & 9:30                    \\
Taylor Swift   & Women in Music Speech\tablefootnote{Women in Music Speech by Taylor Swift:  \url{https://www.youtube.com/watch?v=ZVpkFb9-fts} Last accessed: 07/23}   & 10:00                   \\
Donald Trump     & Victory Speech\tablefootnote{Victory Speech by Donald Trump: \url{https://www.youtube.com/watch?v=Qsvy10D5rtc} Last accessed: 07/23}          & 10:00                   \\ \hline
\textit{\textbf{Total}}                                    &                         & \textit{\textbf{62:22}} \\ \hline
\end{tabular}
\end{table}

\begin{figure}[t]
    \centering
    \includegraphics[scale=1]{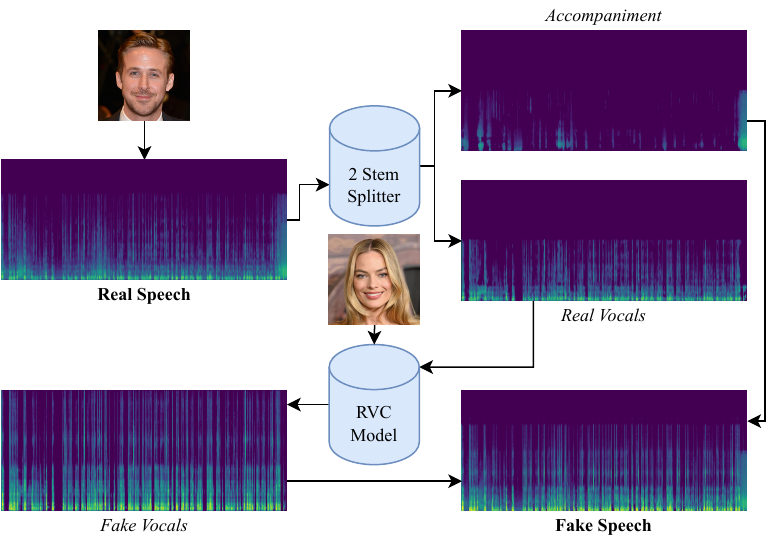}
    \caption{Overview of the Retrieval-based Voice Conversion process to generate DeepFake speech with Ryan Gosling's speech converted to Margot Robbie. Conversion is run on the extracted vocals before being layered on the original background ambience.}
    \label{fig:conversiondiagram}
\end{figure}

An example of voice conversion from real to fake speech can be found in Figure \ref{fig:conversiondiagram}. First, the real speech is split via the two-stem model \cite{hennequin2020spleeter} from Spleeter, which is an encoder-decoder Convolutional Neural Network (CNN) in a U-Net architecture. The model consists of $12$ layers, with $6$ layers each for the encoder and decoder networks. Following splitting of the real vocals and accompaniment tracks, the vocals are then converted using a Retrieval-based Voice Conversion (RVC) model to another individual. Finally, the original accompaniment and the RVC vocals are combined to form a fake speech track. The reason for splitting the tracks is so the style of the deepfake voice is not converted to any background noise, such as audience cheers or laughter. That is, the aim of this approach is to preserve ambient sounds while converting only the speaker's voice. 

The aforementioned style conversion process is performed on each of the audio tracks gathered and cropped from Table \ref{tab:dataset-length}, and features are extracted for every 1-second of audio signal. In total, $26$ features are extracted. Further information about the feature extraction can be found in \cite{mcfee2015librosa}. Those features include the Chromagram, which can be calculated from the Short-Time Fourier Transform $X$ of a signal at frame $m$ and frequency bin $k$:

\begin{equation}
    \text{SG}(m, k) = |X(m, k)|,
\end{equation}

\noindent and then normalising the Chroma Bands which are spectrogram bins into, also known as chroma bands. From this, the Spectral Centroid $SC$ can then be calculated, which is the location of the centre of mass in the spectrum:

\begin{equation}
    \text{SC}(m) = \frac{\sum_{k} k \cdot \text{SG}(m, k)}{\sum_{k} \text{SG}(m, k)}.
\end{equation}

\noindent Similarly, the Spectral Bandwidth $SB$, which is the difference in frequencies around the centroid, is calculated as:

\begin{equation}
    \text{SB}(m) = \sqrt{\frac{\sum_{k} (k - \text{SC}(m))^2 \cdot \text{SG}(m, k)}{\sum_{k} \text{SG}(m, k)}}.
\end{equation}

The Spectral Rolloff (SR) is calculated via the frequency below 85\% of the total spectral energy. Following the spectral features, the Zero Crossing Rate $ZCR$ is measured by observing the rate at which the signal (containing $N$ samples) changes sign via the sign function $sgn$. That is, how often the signal crosses $x=0$ and changes from positive to negative or vice-versa:

\begin{equation}
    \text{ZCR}(m) = \frac{1}{2N}\sum_{n=1}^{N-1} |sgn(x[n]) - sgn(x[n-1])|.
\end{equation}

\noindent The Root Mean Square $RMS$ is then calculated from audio signal $x[n]$ at frame $m$:

\begin{equation}
    \text{RMS}(m) = \sqrt{\frac{1}{N}\sum_{n=0}^{N-1}x[n]^2}.
\end{equation}

\noindent Finally, the first 20 Mel-Frequency Cepstral Coefficients (MFCCs) are calculated. The powers of the Short-Time Fourier Transform are mapped to the Mel-scale by applying a triangular window. Each of the $\log$ is taken from the power spectrum, and a Discrete Cosine Transform (DCT) is applied.

Given that every individual is used to generate seven fake tracks, there is an extreme data imbalance. For this reason, the data is balanced at a 1:1 ratio between real and fake speech by undersampling the data belonging to the fake class. A sample equal to the length of the real data is selected at random and used in the following methodology. 

\subsection{DeepFake Voice Conversion}
The approach selected for converting speech is the RVC model\footnote{Implementation of RVC can be found at: \url{https://github.com/RVC-Project/Retrieval-based-Voice-Conversion-WebUI}}, which is based on the VITS architecture \cite{kim2021conditional}. VITS is an adversarially trained Conditional Variational Autoencoder, originally intended for Text-to-Speech. Pitch estimation from the input speech is determined through the CREPE model \cite{kim2018crepe}, which is a deep CNN model. While the approach achieves competitive state-of-the-art voice conversion results, RVC is also chosen due to its ability to quickly convert short speech samples, leading to the possibility of it being used in real-time. That is, a microphone can be used as an input source to the model, enabling individuals to convert their voice to another's during a call. 

Models for the individuals studied in this work are retrieved from the Huggingface Model Hub\footnote{Information on Huggingface Model Hub can be found at: \url{https://huggingface.co/models}} following a search within the AI Hub Discord server, which provides search functionality\footnote{Information on the AI Hub can be found at: \url{https://discord.me/aihub}}. For diversity, both male and female voices are selected from a range of public figures. Selected for this study are RVC version 2 models trained on the current and former two United States Presidents, Joe Biden (500 epochs), Donald Trump (600 epochs), and Barack Obama (300 epochs). Public figures included Elon Musk (350 epochs) and Linus Sebastian (300 epochs). Singers included Taylor Swift (300 epochs). Finally, two actors were chosen, Ryan Gosling (350 epochs) and Margot Robbie (350 epochs). 

\subsection{Machine Learning Model}
Following the feature extraction from each 1-second block of audio, this study then implements a range of various ML models. The goal of the models is to perform binary classification of the speech, learning whether the audio is speech spoken naturally by a human being, or has been, tampered with by retrieval-based voice conversion. The models are chosen from a range of different statistical approaches for comparison. 

The models trained are: Extreme Gradient Boosting (XGBoost) \cite{chen2016xgboost}, Random Forests \cite{breiman2001random}, Quadratic and Linear Discriminant analyses \cite{fisher1936use}, Ridge Regression \cite{hoerl1970ridge} (linear regression with L2 regularisation), Gaussian and Bernoulli Naive Bayes \cite{bayes1763lii}, K-Nearest Neighbours \cite{fix1951discriminatory}, Support Vector Machines \cite{hearst1998support}, Stochastic Gradient Descent \cite{robbins1951stochastic}, and Gaussian Process \cite{williams1995gaussian}. Along with classification metrics, inference time is also considered, since detection of AI-generated speech may be useful in real-time during a conference or phone call. The average inference time is calculated by measuring and averaging the inference time of 1000 random data objects within the dataset. Towards hyperparameter optimisation, the XGBoost and Random Forests are optimised through a linear search of $\{10,20,30,...,500\}$ boosting rounds and forest sizes, respectively. The K-Nearest Neighbor problem space is searched in a similar fashion, with the neighbour ensembles being sized $\{1,2,3,...,100\}$.

Along with classical accuracy, further metrics are also considered for model comparison. These are precision, which measures the rate at which predicted positives are correct among all positive predictions, and allows for false-positive analysis:

\begin{equation}
\text{Precision} = \frac{\text{True positives}}{\text{True positives + False positives}}.
\end{equation}

\noindent Precision is important since a high precision would minimise false accusations of AI-generated speech when the audio is, in fact, natural voice. Similarly, recall which is a measure of how many positive cases are correctly predicted, which enables analysis of false-negative predictions:

\begin{equation}
    \text{Recall} = \frac{\text{True positives}}{\text{True positives + False negatives}}.
\end{equation}

\noindent Higher recall suggests that the model is not falsely classifying AI-generated speech as human speech. These results are then combined to compute the F-1 score:

\begin{equation}
    \text{F1 score} = 2 \times \frac{\text{Precision} \times \text{Recall}}{\text{Precision} + \text{Recall}}.
\end{equation}

\noindent The Matthews Correlation Coefficient (MCC) is then considered, which is a metric that considers all potential correct and incorrect predictions, calculated as:

\begin{equation}
\text{MCC} = \frac{\text{T} \times \text{TN} - \text{FP} \times \text{FN}}{\sqrt{(\text{TP} + \text{FP}) \times (\text{TP} + \text{FN}) \times (\text{TN} + \text{FP}) \times (\text{TN} + \text{FN})}}.
\end{equation}

\noindent MCC is measured in the range of -1 to 1. -1 is the complete disagreement between predictions and labels, +1 is a perfect classifier, and an MCC of 0 suggests random predictions. 

Finally, the Receiver Operating Characteristic Area Under the Curve (ROC AUC) is calculated. The ROC AUC is important for probabilistic approaches to class prediction, since it is a test of predictive ability across probability thresholds. The ROC is a plot of recall and false positives given these thresholds, and the AUC is the measurement of the area under the plotted curve. The metric is ranged from $0$ to $1$, where $0.5$ is a random classifier and $1$ is perfect. 

Each model is trained over 10-fold cross validation, with data splits set for replicability and direct comparison via a random seed of $42$. 

\subsection{Experimental Hardware and Software}
All experiments in this study were trained and executed on an Intel Core i7 CPU with a clock speed of 3.7GHz. Voice conversion was executed on a GPU due to the use of the CREPE algorithm, and an Nvidia RTX 2080Ti was used (4,352 CUDA cores). The features were extracted from audio using the Librosa\cite{mcfee2015librosa} library, and ML models were implemented with scikit-learn\cite{scikit-learn}.

\section{Results and Observations}
\label{sec:results}
This section contains the observations made and results found within the experiments. 

\subsection{Dataset Analysis}

\begin{table}[]
\caption{Observed statistics in the dataset between the two classes of data. }
\label{tab:stats}
\footnotesize
\centering
\begin{tabular}{@{}lrrrrrr@{}}
\toprule
\multirow{2}{*}{\textbf{Attribute}}                                                      & \multicolumn{3}{l}{\textbf{Real}}                                                                                                    & \multicolumn{3}{l}{\textbf{Fake}}                                                                                                    \\ \cmidrule(l){2-7} 
                                                                                & \multicolumn{1}{l}{\textit{\textbf{Mean}}} & \multicolumn{1}{l}{\textit{\textbf{Med.}}} & \multicolumn{1}{l}{\textit{\textbf{Std.}}} & \multicolumn{1}{l}{\textit{\textbf{Mean}}} & \multicolumn{1}{l}{\textit{\textbf{Med.}}} & \multicolumn{1}{l}{\textit{\textbf{Std.}}} \\ \cmidrule(r){1-1}
\textit{\textbf{Chromagram}}                                                    & 0.41                                       & 0.40                                       & \multicolumn{1}{r|}{0.07}                  & 0.43                                       & 0.43                                       & 0.07                                       \\
\textit{\textbf{\begin{tabular}[c]{@{}l@{}}Root Mean \\ Square\end{tabular}}}   & 0.04                                       & 0.03                                       & \multicolumn{1}{r|}{0.03}                  & 0.04                                       & 0.03                                       & 0.02                                       \\
\textit{\textbf{\begin{tabular}[c]{@{}l@{}}Spectral \\ Centroid\end{tabular}}}  & 2541.34                                    & 2353.97                                    & \multicolumn{1}{r|}{1,253.73}              & 2897.06                                    & 2,762.58                                   & 800.40                                     \\
\textit{\textbf{\begin{tabular}[c]{@{}l@{}}Spectral \\ Bandwidth\end{tabular}}} & 2883.99                                    & 2806.94                                    & \multicolumn{1}{r|}{1,080.81}              & 3216.61                                    & 3193.74                                    & 545.97                                     \\
\textit{\textbf{Rolloff}}                                                       & 4683.44                                    & 4094.33                                    & \multicolumn{1}{r|}{2602.42}               & 5271.79                                    & 5061.05                                    & 1572.49                                    \\
\textit{\textbf{\begin{tabular}[c]{@{}l@{}}Zero \\ Crossing Rate\end{tabular}}} & 0.06                                       & 0.05                                       & \multicolumn{1}{r|}{0.04}                  & 0.08                                       & 0.07                                       & 0.03                                       \\
\textit{\textbf{MFCC 1}}                                                        & -376.65                                    & -354.28                                    & \multicolumn{1}{r|}{84.12}                 & -388.48                                    & -378.02                                    & 74.31                                      \\
\textit{\textbf{MFCC 2}}                                                        & 158.25                                     & 162.36                                     & \multicolumn{1}{r|}{40.18}                 & 131.86                                     & 132.94                                     & 25.62                                      \\
\textit{\textbf{MFCC 3}}                                                        & -29.60                                     & -24.75                                     & \multicolumn{1}{r|}{31.77}                 & -19.80                                     & -16.72                                     & 21.92                                      \\
\textit{\textbf{MFCC 4}}                                                        & 14.66                                      & 11.63                                      & \multicolumn{1}{r|}{23.63}                 & 27.96                                      & 27.38                                      & 19.07                                      \\
\textit{\textbf{MFCC 5}}                                                        & -6.33                                      & -11.72                                     & \multicolumn{1}{r|}{23.88}                 & -6.31                                      & -5.15                                      & 15.61                                      \\
\textit{\textbf{MFCC 6}}                                                        & 3.40                                       & 5.17                                       & \multicolumn{1}{r|}{15.01}                 & 11.41                                      & 12.48                                      & 12.54                                      \\
\textit{\textbf{MFCC 7}}                                                        & -8.61                                      & -7.29                                      & \multicolumn{1}{r|}{12.31}                 & -10.37                                     & -11.01                                     & 10.49                                      \\
\textit{\textbf{MFCC 8}}                                                        & -7.40                                      & -6.79                                      & \multicolumn{1}{r|}{10.82}                 & -4.73                                      & -4.59                                      & 7.23                                       \\
\textit{\textbf{MFCC 9}}                                                        & -8.41                                      & -8.28                                      & \multicolumn{1}{r|}{11.28}                 & -3.48                                      & -4.31                                      & 8.04                                       \\
\textit{\textbf{MFCC 10}}                                                       & -10.98                                     & -13.21                                     & \multicolumn{1}{r|}{8.74}                  & -7.26                                      & -6.84                                      & 8.81                                       \\
\textit{\textbf{MFCC 11}}                                                       & -3.27                                      & -3.20                                      & \multicolumn{1}{r|}{7.53}                  & -1.21                                      & -1.72                                      & 7.78                                       \\
\textit{\textbf{MFCC 12}}                                                       & -5.88                                      & -5.45                                      & \multicolumn{1}{r|}{6.95}                  & -3.00                                      & -2.94                                      & 5.93                                       \\
\textit{\textbf{MFCC 13}}                                                       & -2.07                                      & -1.99                                      & \multicolumn{1}{r|}{4.52}                  & -1.25                                      & -0.90                                      & 5.63                                       \\
\textit{\textbf{MFCC 14}}                                                       & -1.67                                      & -1.64                                      & \multicolumn{1}{r|}{5.27}                  & -2.55                                      & -2.88                                      & 5.39                                       \\
\textit{\textbf{MFCC 15}}                                                       & -3.06                                      & -2.88                                      & \multicolumn{1}{r|}{4.80}                  & -2.16                                      & -1.93                                      & 4.97                                       \\
\textit{\textbf{MFCC 16}}                                                       & -2.29                                      & -1.07                                      & \multicolumn{1}{r|}{5.59}                  & -0.99                                      & -0.62                                      & 5.59                                       \\
\textit{\textbf{MFCC 17}}                                                       & -3.09                                      & -2.95                                      & \multicolumn{1}{r|}{4.53}                  & -3.55                                      & -3.48                                      & 4.65                                       \\
\textit{\textbf{MFCC 18}}                                                       & -4.85                                      & -4.05                                      & \multicolumn{1}{r|}{4.91}                  & -1.38                                      & -1.81                                      & 4.40                                       \\
\textit{\textbf{MFCC 19}}                                                       & -1.79                                      & -1.99                                      & \multicolumn{1}{r|}{4.40}                  & -3.72                                      & -3.41                                      & 5.29                                       \\
\textit{\textbf{MFCC 20}}                                                       & -4.08                                      & -3.29                                      & \multicolumn{1}{r|}{6.23}                  & -4.77                                      & -4.22                                      & 4.59                                       \\ \bottomrule
\end{tabular}
\end{table}

Prior to implementation of Machine Learning, Table \ref{tab:stats} shows the observed statistics between each the real and fake datasets. It can be seen that there are large differences between some of the attributes, for example, the mean Spectral Centroid of the fake data is 2897.06 whereas is much lower for the real data at 2541.34. The mean value of the first and second MFCCs also substantially different values given the classes. On the other hand, several feature sets exhibit similarity, such as MFCC 5. 

\begin{table}[]
\caption{Results of the unpaired t-test for each attribute between the real and fake classes of data. }
\label{tab:t-test}
\footnotesize
\centering
\begin{tabular}{@{}lrrc@{}}
\toprule
\textbf{Attribute}                                                              & \multicolumn{1}{l}{\textbf{T-Statistic}} & \multicolumn{1}{l}{\textbf{P-Value}} & \textbf{Significance?} \\ \midrule
\textit{\textbf{Chromagram}}                                                    & -17.488                                  & 1.25E-67                             & Y                      \\
\textit{\textbf{\begin{tabular}[c]{@{}l@{}}Root Mean \\ Square\end{tabular}}}   & 7.799                                    & 6.78E-15                             & Y                      \\
\textit{\textbf{\begin{tabular}[c]{@{}l@{}}Spectral \\ Centroid\end{tabular}}}  & -18.351                                  & 3.52E-74                             & Y                      \\
\textit{\textbf{\begin{tabular}[c]{@{}l@{}}Spectral \\ Bandwidth\end{tabular}}} & -21.078                                  & 7.70E-97                             & Y                      \\
\textit{\textbf{Rolloff}}                                                       & -14.848                                  & 2.02E-49                             & Y                      \\
\textit{\textbf{\begin{tabular}[c]{@{}l@{}}Zero \\ Crossing Rate\end{tabular}}} & -17.173                                  & 2.65E-65                             & Y                      \\
\textit{\textbf{MFCC 1}}                                                        & 8.087                                    & 6.69E-16                             & Y                      \\
\textit{\textbf{MFCC 2}}                                                        & 42.5                                     & 0.00E+00                             & Y                      \\
\textit{\textbf{MFCC 3}}                                                        & -19.467                                  & 4.24E-83                             & Y                      \\
\textit{\textbf{MFCC 4}}                                                        & -33.626                                  & 8.95E-237                            & Y                      \\
\textit{\textbf{MFCC 5}}                                                        & -0.05                                    & 9.61E-01                             & N                      \\
\textit{\textbf{MFCC 6}}                                                        & -31.418                                  & 3.89E-208                            & Y                      \\
\textit{\textbf{MFCC 7}}                                                        & 8.349                                    & 7.65E-17                             & Y                      \\
\textit{\textbf{MFCC 8}}                                                        & -15.774                                  & 1.74E-55                             & Y                      \\
\textit{\textbf{MFCC 9}}                                                        & -27.323                                  & 2.01E-159                            & Y                      \\
\textit{\textbf{MFCC 10}}                                                       & -23.012                                  & 1.17E-114                            & Y                      \\
\textit{\textbf{MFCC 11}}                                                       & -14.627                                  & 4.98E-48                             & Y                      \\
\textit{\textbf{MFCC 12}}                                                       & -24.232                                  & 1.25E-126                            & Y                      \\
\textit{\textbf{MFCC 13}}                                                       & -8.665                                   & 5.08E-18                             & Y                      \\
\textit{\textbf{MFCC 14}}                                                       & 8.989                                    & 2.88E-19                             & Y                      \\
\textit{\textbf{MFCC 15}}                                                       & -9.949                                   & 3.14E-23                             & Y                      \\
\textit{\textbf{MFCC 16}}                                                       & -12.613                                  & 3.08E-36                             & Y                      \\
\textit{\textbf{MFCC 17}}                                                       & 5.345                                    & 9.22E-08                             & Y                      \\
\textit{\textbf{MFCC 18}}                                                       & -40.388                                  & 0.00E+00                             & Y                      \\
\textit{\textbf{MFCC 19}}                                                       & 21.553                                   & 4.37E-101                            & Y                      \\
\textit{\textbf{MFCC 20}}                                                       & 6.894                                    & 5.69E-12                             & Y                      \\ \bottomrule
\end{tabular}
\end{table}

Subsequently, Table \ref{tab:t-test} then shows the results for the unpaired t-test between datasets. It can be observed that all features with exception of the aforementioned 5$^{th}$ Mel-Frequency Cepstral Coefficient showed statistical significance between the two classes of data. Given that most features have significantly different means between the two classes, they are potentially valuable to distinguish and thus useful to train machine learning models. 

\begin{figure}[t]
    \centering
    \includegraphics[scale=0.57]{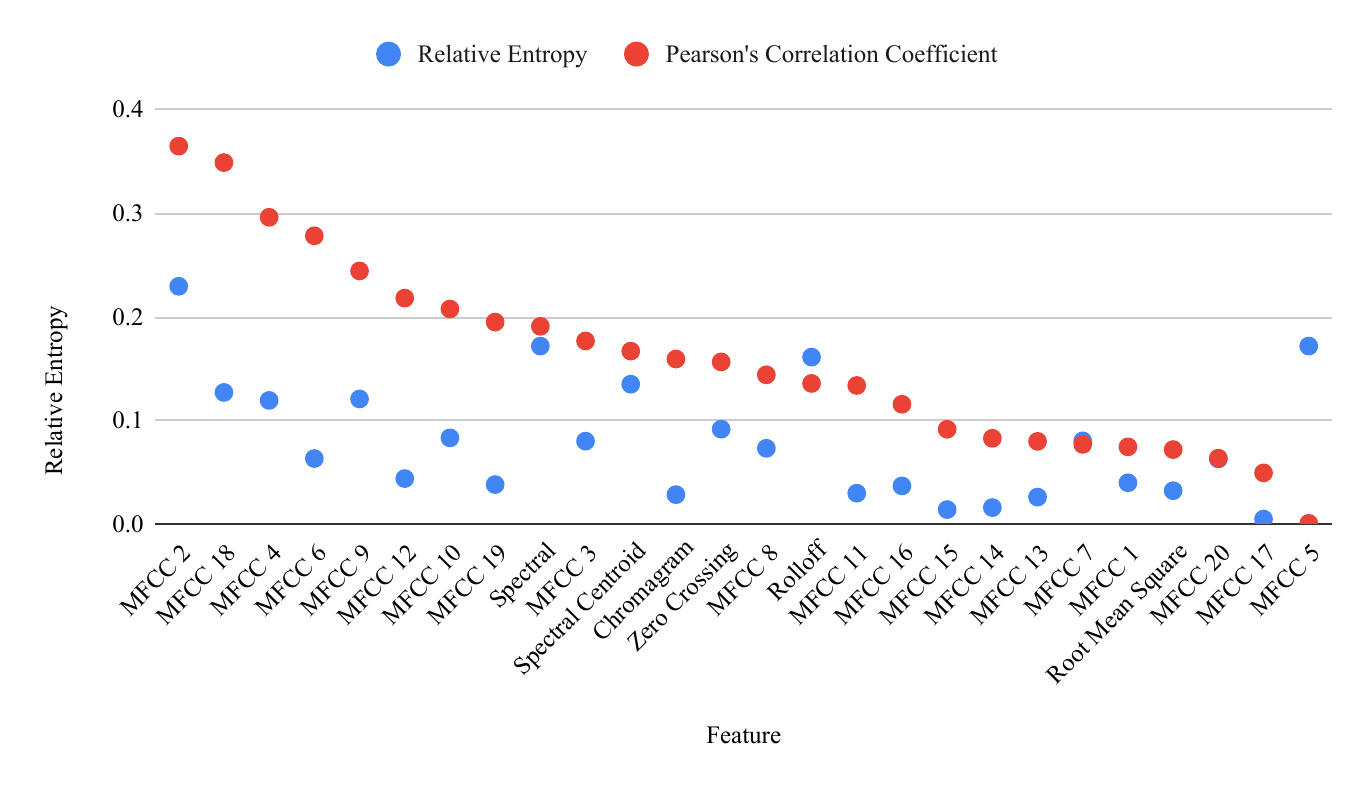}
    \caption{Pearson's Correlation Coefficient and Relative entropy for all of the extracted features when used for binary classification of real or AI-generated vocals (sorted by Pearson's).}
    \label{fig:entropygraph}
\end{figure}

Figure \ref{fig:entropygraph} shows the correlation coefficients and relative entropy of each of the extracted features when used for binary classification of real or AI-generated vocals. The highest correlation between the attributes and class value was observed to be the 2$^{nd}$ MFCC, with a coefficient of 0.36. This was closely followed by the 18$^{th}$ at a value of 0.35. The least correlation between attribute and class value were the 20$^{th}$, 17$^{th}$, and 5$^{th}$ MFCCs, which had correlation coefficients of 0.06, 0.05, and 0.0005, respectively. The highest relative entropy feature was the 2$^{nd}$ MFCC with a value of 0.23, which was followed by the 5$^{th}$ at 0.172. The top feature that was not an MFCC came third, the Spectral Bandwidth value with a relative entropy of 0.172. Towards the lowest end of the values came the 15$^{th}$ and 17$^{th}$ MFCCs, which had respective values of 0.014 and 0.005. It is interesting to note that the lowest correlation coefficient feature was also that which had no statistical significance between the two classes of data, however, was noted to have relatively high information gain.  

\begin{table}[t]
\caption{Metrics when using a single rule-based classifier to split predictions via the 2$^{nd}$ Mel Frequency Cepstral Coefficient. Overall, using this feature, a mean accuracy of 69.84\% was observed over 10-fold cross validation.}
\label{tab:single-rule}
\centering
\begin{tabular}{llllll}
\hline
\multirow{2}{*}{\textbf{Class}}    & \multicolumn{5}{l}{\textbf{Metric}}                                                                                                 \\ \cline{2-6} 
                                   & \textit{\textbf{Precision}} & \textit{\textbf{Recall}} & \textit{\textbf{F1-Score}} & \textit{\textbf{MCC}} & \textit{\textbf{ROC}} \\ \cline{1-1}
\textit{\textbf{Real}}             & 0.708                       & 0.677                    & 0.692                      & 0.397                 & 0.698                 \\
\textit{\textbf{Fake}}             & 0.690                       & 0.720                    & 0.705                      & 0.397                 & 0.698                 \\
\textit{\textbf{Weighted Average}} & 0.699                       & 0.698                    & 0.698                      & 0.397                 & 0.698                 \\ \hline
\end{tabular}
\end{table}

\begin{figure}[]
    \centering
    \includegraphics[scale=0.65]{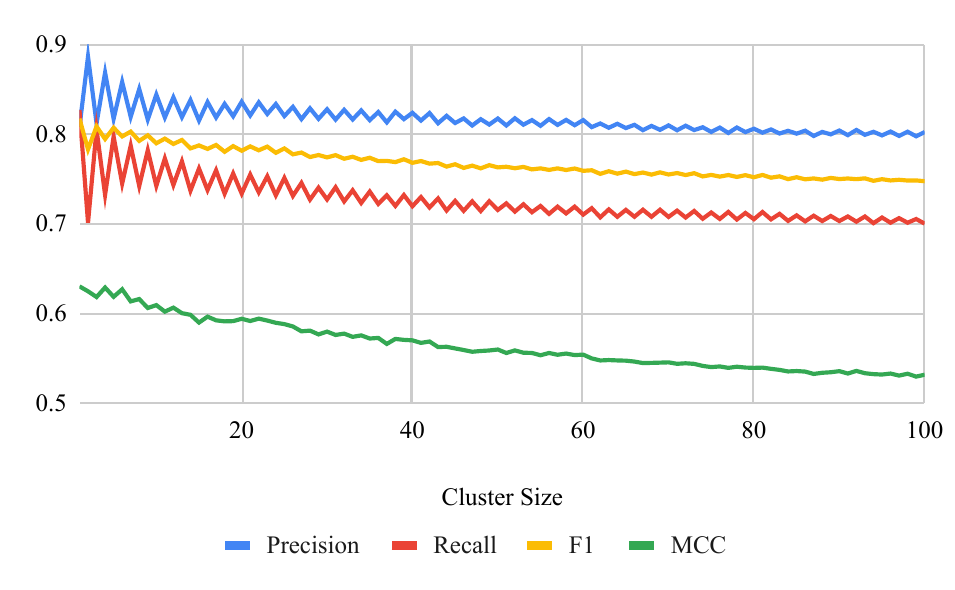}
    \caption{Results for the KNN model when searching for the most optimal cluster.}
    \label{fig:knn-results}
\end{figure}

\begin{figure}[]
    \centering
    \includegraphics[scale=0.65]{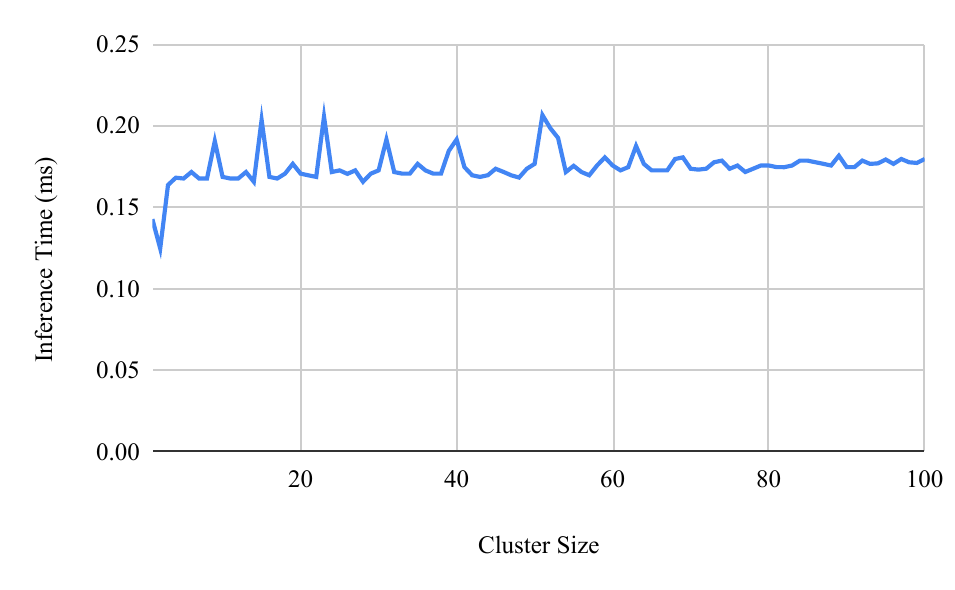}
    \caption{Observed average inference time for the KNN models to classify 1-second of audio data.}
    \label{fig:knn-inference}
\end{figure}

\begin{figure}[]
    \centering
    \includegraphics[scale=0.65]{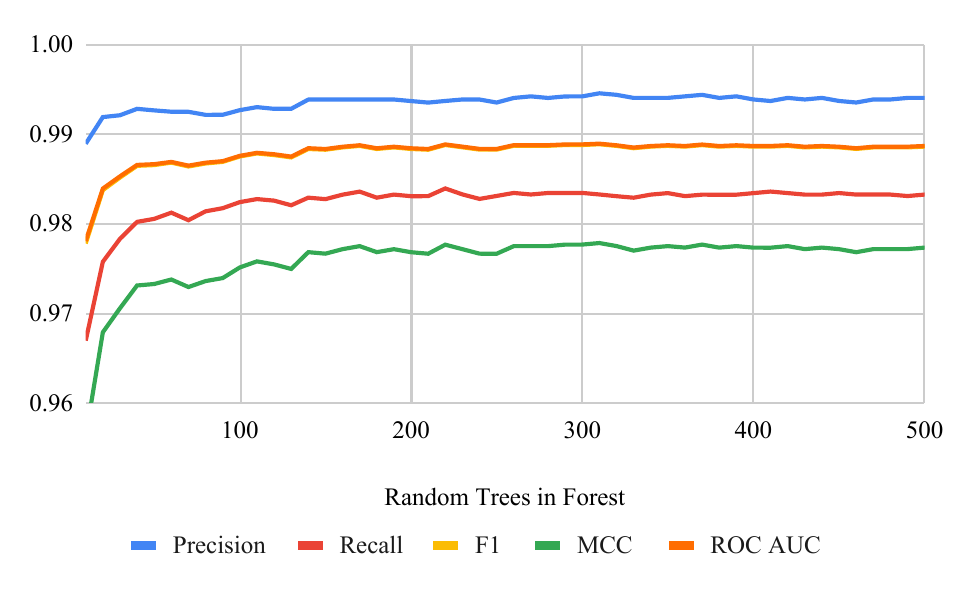}
    \caption{Results for the Random Forest model when searching for the most optimal ensemble size.}
    \label{fig:rf-results}
\end{figure}

\begin{figure}[]
    \centering
    \includegraphics[scale=0.65]{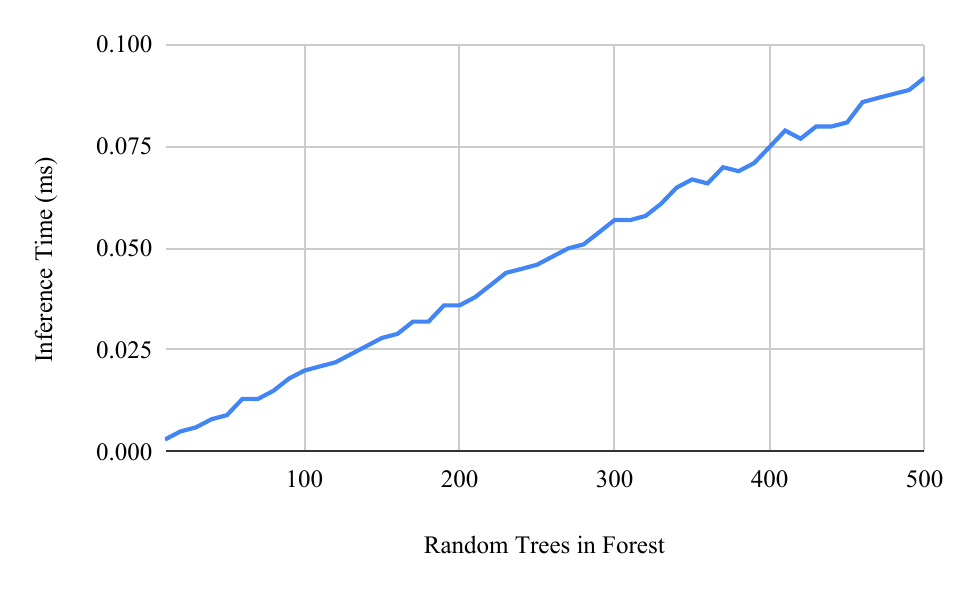}
    \caption{Observed average inference time for the Random Forest models to classify 1-second of audio data.}
    \label{fig:rf-inference}
\end{figure}

Furthermore, Table \ref{tab:single-rule} shows observed metrics when making a prediction by splitting values based on the feature with the most correlation, the 2$^{nd}$ Mel-frequency Cepstral Coefficient, which had a Pearson's correlation coefficient of 0.36. The analysis shows that by splitting the data on this one attribute, a mean accuracy of 69.84\% can be achieved. 

\subsection{Hyperparameter Optimisation}
This section details the findings within linear hyperparameter optimisation for the clustering, random forest ensemble, and XGBoost models. 

Figure \ref{fig:knn-results} shows the results for the KNN models. Interestingly, the highest performing model was the smallest cluster of 1 nearest neighbour, which scored 81.48\% accuracy, 0.827 recall, 0.817 F-Score, an MCC of 0.63, and an area under the ROC Curve of 0.815. However, the highest precision value was 0.886 when using two nearest neighbours as the predictors. As can be observed in Figure \ref{fig:knn-inference}, this model took, on average, 0.143 milliseconds to classify 1-second of audio data as real or fake. 

Unlike the KNN, the Random Forest results were moreso relative to one another given a number of trees in the forest. Figure \ref{fig:rf-results} shows the classification results for each ensemble size. As can be observed, the model containing 310 trees scored an average 98.89\% accuracy over the 10-folds of data. Similarly, the recall, precision, F-Score, MCC, and ROC area were 0.995, 0.983, 0.989, and 0.989, respectively. Figure \ref{fig:rf-inference} shows the inference time given an ensemble size. As could be expected, inference time increases relatively linearly given the ensemble size, which is shown in Figure \ref{fig:rf-inference}. The aforementioned ensemble of 310 random trees took an average of 0.057 milliseconds to classify 1-second of audio data. 

\begin{figure}[t]
    \centering
    \includegraphics[scale=0.65]{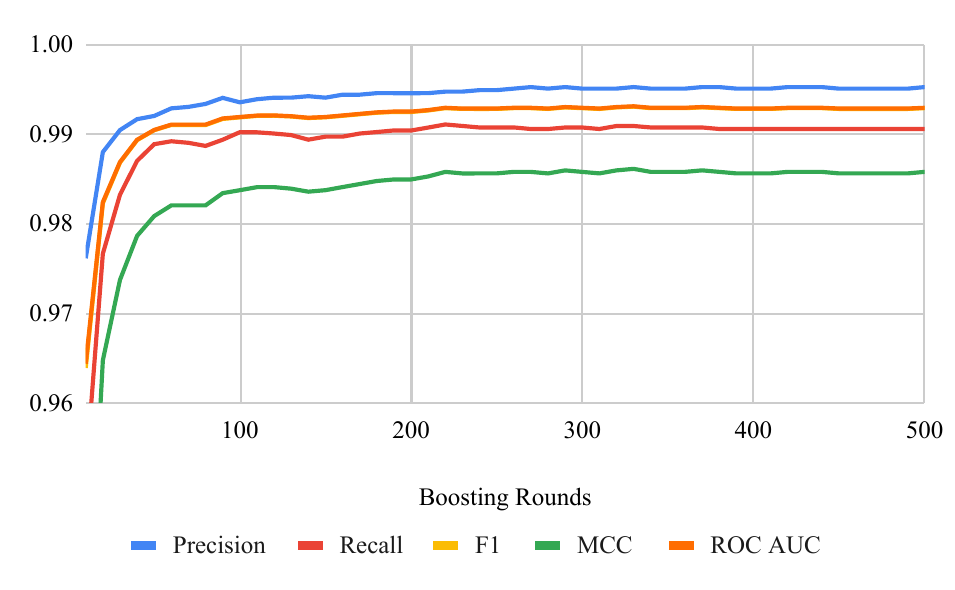}
    \caption{Results for the XGBoost model when searching for the most optimal number of boosting rounds.}
    \label{fig:xg-results}
\end{figure}

\begin{figure}[t]
    \centering
    \includegraphics[scale=0.65]{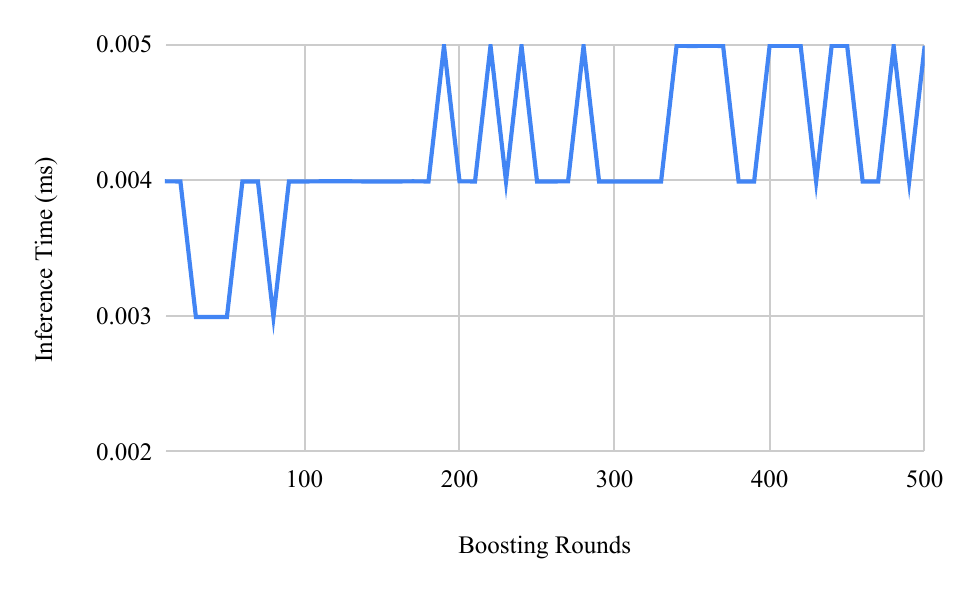}
    \caption{Observed average inference time for the XGBoost models to classify 1-second of audio data.}
    \label{fig:xg-inference}
\end{figure}

Figure \ref{fig:xg-results} shows the affect of the number of boosting rounds for the XGBoost model when classifying the data. A dramatic increase of around 3\% accuracy gain can be observed between 10 to 50 boosting rounds, before the results become relatively stagnant. The overall best-performing approach was when following 330 rounds of boosting, which led to a classification accuracy of 99.3\%. The precision of this model was 0.995, the recall was 0.991, with an F-Score of 0.993. The MCC was observed at 0.986 with an area under the ROC curve of 0.993. Figure \ref{fig:xg-inference} shows the inference results, which, within a range of 0.001 milliseconds, were the same. The XGBoost model following 330 rounds took an average of 0.004 milliseconds to predict the class belonging to 1-second of audio data.

\subsection{Results Comparison}
This section compares all of the results for the models trained in this study. As described previously, 10-fold cross validation is used and the average metrics along with variance are reported. 

\begin{table}[t]
\centering
\caption{Comparison of averaged validation metrics over 10-fold cross validation for the machine learning models. Inference time denotes the average time taken for the model to predict the class of 1-second of audible speech.}
\label{tab:results-comparison-avg}
\footnotesize
\begin{tabular}{@{}lrrrrrrr@{}}
\toprule
\multirow{2}{*}{Model}                                                                          & \multicolumn{7}{l}{\textbf{Mean Value over 10-fold Cross Validation}}                                                                                                                                                                                                                                                                                                               \\ \cmidrule(l){2-8} 
                                                                                                & \multicolumn{1}{l}{\textit{\textbf{Acc.}}} & \multicolumn{1}{l}{\textit{\textbf{Prec.}}} & \multicolumn{1}{l}{\textit{\textbf{Rec.}}} & \multicolumn{1}{l}{\textit{\textbf{F1}}} & \multicolumn{1}{l}{\textit{\textbf{MCC}}} & \multicolumn{1}{l}{\textit{\textbf{ROC AUC}}} & \multicolumn{1}{l}{\textit{\textbf{\begin{tabular}[c]{@{}l@{}}Inference \\ Time (ms)\end{tabular}}}} \\ \cmidrule(r){1-1}
\textit{\textbf{XGBoost (330)}}                                                                 & 0.993                                      & 0.995                                       & 0.991                                      & 0.993                                    & 0.986                                     & 0.993                                         & 0.004                                                                                                \\
\textit{\textbf{Random Forest (310)}}                                                           & 0.989                                      & 0.995                                       & 0.983                                      & 0.989                                    & 0.978                                     & 0.989                                         & 0.057                                                                                                \\
\textit{\textbf{\begin{tabular}[c]{@{}l@{}}Quadratic \\ Discriminant \\ Analysis\end{tabular}}} & 0.948                                      & 0.969                                       & 0.924                                      & 0.946                                    & 0.896                                     & 0.948                                         & 0.002                                                                                                \\
\textit{\textbf{\begin{tabular}[c]{@{}l@{}}Linear \\ Discriminant \\ Analysis\end{tabular}}}    & 0.889                                      & 0.886                                       & 0.893                                      & 0.889                                    & 0.778                                     & 0.889                                         & 0.001                                                                                                \\
\textit{\textbf{Ridge}}                                                                         & 0.883                                      & 0.884                                       & 0.882                                      & 0.883                                    & 0.767                                     & 0.883                                         & 0.001                                                                                                \\
\textit{\textbf{\begin{tabular}[c]{@{}l@{}}Naïve Bayes \\ (Gaussian)\end{tabular}}}             & 0.830                                      & 0.864                                       & 0.784                                      & 0.822                                    & 0.664                                     & 0.830                                         & 0.001                                                                                                \\
\textit{\textbf{KNN (1)}}                                                                       & 0.815                                      & 0.808                                       & 0.827                                      & 0.817                                    & 0.630                                     & 0.815                                         & 0.143                                                                                                \\
\textit{\textbf{SVM}}                                                                           & 0.723                                      & 0.815                                       & 0.576                                      & 0.675                                    & 0.465                                     & 0.723                                         & 0.605                                                                                                \\
\textit{\textbf{\begin{tabular}[c]{@{}l@{}}Naïve Bayes \\ (Bernoulli)\end{tabular}}}            & 0.692                                      & 0.742                                       & 0.587                                      & 0.655                                    & 0.391                                     & 0.691                                         & 0.001                                                                                                \\
\textit{\textbf{\begin{tabular}[c]{@{}l@{}}Stochastic \\ Gradient Descent\end{tabular}}}        & 0.668                                      & 0.732                                       & 0.760                                      & 0.681                                    & 0.407                                     & 0.673                                         & 0.001                                                                                                \\
\textit{\textbf{Gaussian Process}}                                                              & 0.614                                      & 0.997                                       & 0.229                                      & 0.372                                    & 0.358                                     & 0.614                                         & 0.561                                                                                                \\ \bottomrule
\end{tabular}%
\end{table}

Table \ref{tab:results-comparison-avg} shows the overall comparison of results for each of the machine learning models. As can be observed, the best-performing model was the Extreme Gradient Boosting model, which scored an average 99.3\% accuracy over the 10-folds of data. However, the third-best performing model, Quadratic Discriminant Analysis, may have scored 94.8\% accuracy, but could classify audio data in half the time at 0.002 milliseconds per object. The performance of extreme gradient boosting and random forests compared to the other models suggests that an ensemble approach is most useful for this classification problem. 

\begin{table}[t]
\centering
\caption{Comparison of standard deviation within the classification metrics over 10-fold cross validation for the machine learning models.}
\label{tab:results-comparison-std}
\footnotesize
\begin{tabular}{@{}lrrrrrr@{}}
\toprule
\multirow{2}{*}{Model}                                                                          & \multicolumn{6}{l}{\textbf{Mean Value over 10-fold Cross Validation}}                                                                                                                                                                                                        \\ \cmidrule(l){2-7} 
                                                                                                & \multicolumn{1}{l}{\textit{\textbf{Acc.}}} & \multicolumn{1}{l}{\textit{\textbf{Prec.}}} & \multicolumn{1}{l}{\textit{\textbf{Rec.}}} & \multicolumn{1}{l}{\textit{\textbf{F1}}} & \multicolumn{1}{l}{\textit{\textbf{MCC}}} & \multicolumn{1}{l}{\textit{\textbf{ROC AUC}}} \\ \cmidrule(r){1-1}
\textit{\textbf{XGBoost (330)}}                                                                 & 0.002                                      & 0.002                                       & 0.005                                      & 0.002                                    & 0.005                                     & 0.002                                         \\
\textit{\textbf{Random Forest (310)}}                                                           & 0.003                                      & 0.003                                       & 0.005                                      & 0.003                                    & 0.006                                     & 0.003                                         \\
\textit{\textbf{\begin{tabular}[c]{@{}l@{}}Quadratic \\ Discriminant \\ Analysis\end{tabular}}} & 0.004                                      & 0.006                                       & 0.008                                      & 0.005                                    & 0.009                                     & 0.005                                         \\
\textit{\textbf{\begin{tabular}[c]{@{}l@{}}Linear \\ Discriminant \\ Analysis\end{tabular}}}    & 0.009                                      & 0.013                                       & 0.009                                      & 0.009                                    & 0.018                                     & 0.009                                         \\
\textit{\textbf{Ridge}}                                                                         & 0.008                                      & 0.011                                       & 0.008                                      & 0.008                                    & 0.016                                     & 0.008                                         \\
\textit{\textbf{\begin{tabular}[c]{@{}l@{}}Naïve Bayes \\ (Gaussian)\end{tabular}}}             & 0.007                                      & 0.013                                       & 0.014                                      & 0.011                                    & 0.015                                     & 0.008                                         \\
\textit{\textbf{KNN (1)}}                                                                       & 0.009                                      & 0.017                                       & 0.011                                      & 0.011                                    & 0.019                                     & 0.009                                         \\
\textit{\textbf{SVM}}                                                                           & 0.012                                      & 0.022                                       & 0.019                                      & 0.018                                    & 0.026                                     & 0.013                                         \\
\textit{\textbf{\begin{tabular}[c]{@{}l@{}}Naïve Bayes \\ (Bernoulli)\end{tabular}}}            & 0.012                                      & 0.014                                       & 0.021                                      & 0.018                                    & 0.023                                     & 0.012                                         \\
\textit{\textbf{\begin{tabular}[c]{@{}l@{}}Stochastic \\ Gradient Descent\end{tabular}}}        & 0.086                                      & 0.178                                       & 0.287                                      & 0.139                                    & 0.146                                     & 0.086                                         \\
\textit{\textbf{Gaussian Process}}                                                              & 0.008                                      & 0.004                                       & 0.013                                      & 0.017                                    & 0.010                                     & 0.006                                         \\ \bottomrule
\end{tabular}%
\end{table}

The variance of the metrics over the 10 folds of data can be found in Table \ref{tab:results-comparison-std}. The two best performing models, XGBoost and Random Forest, also have a low standard deviation. These low values suggest that the models are performing well consistently across the different folds of data during cross-validation. It is worth noting that the QDA approach, while performing slightly worse, is more interpretable and could provide increased explainability along with lower inference time.

\section{Conclusion and Future Work}
\label{sec:conclusion}
This study has addressed some of the growing security implications with generative AI, specifically, those surrounding spoofing with AI-generated human speech. Given the rapidly increasing quality of these systems, it is important that systems should provide transparency in real-time as to the legitimacy of a human voice within a conference or phonecall. AI-generated speech could be used for nefarious purposes, such as impersonation within social engineering attacks. The contributions of this work are comprised of an original audio classification dataset which is released for future work, comprehensive analysis of the statistical significance of audio features extracted from real and AI-generated speech, and finally the optimisation of machine learning models which can predict the legitimacy of speech in real-time. 

To conclude, the remarkable ability of XGBoost and Random Forest models to generalise over folded cross-validation show that it is indeed possible to detect AI-generated speech even with the most state-of-the-art models at the time of writing. Further validation metrics show model robustness, and they were observed to infer data in 0.004 to 0.057 milliseconds, effectively making them real-time classifiers. Current state-of-the-art approaches, such as those discussed in the literature review, are far more computationally expensive and thus may not be able to classify speech in real-time. That is, our approach can detect attacks while they are occurring. 

In future, further improvements to the approach could be made by exploring ensembles of the best performing models. This could enable better generalisation through recognition and correction of model mistakes. The approaches explored may also be improved with additional audio feature representations, increasing the input dimensions and providing more opportunities for rule generation. Finally, the DEEP-VOICE dataset could be expanded with more speakers in the future to further diversify the data and increase generalisation, as well as with the use of different neural speech generation approaches in addition to RVC. 

To finally conclude, this study has proposed a robust machine learning approach to the detection of RVC-based voice conversion, enabling the recognition of voice-based social engineering attacks in real-time. The DEEP-VOICE dataset generated for this study is released to the public to promote interdisciplinary research in AI-generated speech analysis. As the field continues to advance at a rapid pace, proactive approaches are necessary to ensure transparency and foster ethical use of generative AI.

\section{Data Availability Statement}
The datasets generated during and/or analysed during the current study are available in the DEEP-VOICE repository, \url{https://www.kaggle.com/datasets/birdy654/deep-voice-deepfake-voice-recognition}.

\bibliography{bibliography}

\begin{thebibliography}{10}

\bibitem{juefei2022countering}
F.~Juefei-Xu, R.~Wang, Y.~Huang, Q.~Guo, L.~Ma, and Y.~Liu, ``Countering
  malicious deepfakes: Survey, battleground, and horizon,'' {\em International
  journal of computer vision}, vol.~130, no.~7, pp.~1678--1734, 2022.

\bibitem{banks2018deepfakes}
A.~Banks, ``What are deepfakes \& why the future of porn is terrifying,'' {\em
  Highsnobiety. Retrieved}, vol.~17, p.~2021, 2018.

\bibitem{waldrop2020synthetic}
M.~Waldrop, ``Synthetic media: The real trouble with deepfakes,'' {\em Knowable
  Magazine}, vol.~3, 2020.

\bibitem{borel2018clicks}
B.~Borel, ``Clicks, lies and videotape,'' {\em Scientific American}, vol.~319,
  no.~4, pp.~38--43, 2018.

\bibitem{truby2021human}
J.~Truby and R.~Brown, ``Human digital thought clones: the holy grail of
  artificial intelligence for big data,'' {\em Information \& Communications
  Technology Law}, vol.~30, no.~2, pp.~140--168, 2021.

\bibitem{beard2001clones}
J.~J. Beard, ``Clones, bones and twilight zones: protecting the digital persona
  of the quick, the dead and the imaginary,'' {\em J. Copyright Soc'y USA},
  vol.~49, p.~441, 2001.

\bibitem{meese2015posthumous}
J.~Meese, B.~Nansen, T.~Kohn, M.~Arnold, and M.~Gibbs, ``Posthumous personhood
  and the affordances of digital media,'' {\em Mortality}, vol.~20, no.~4,
  pp.~408--420, 2015.

\bibitem{agarwal2019protecting}
S.~Agarwal, H.~Farid, Y.~Gu, M.~He, K.~Nagano, and H.~Li, ``Protecting world
  leaders against deep fakes.,'' in {\em CVPR workshops}, vol.~1, p.~38, 2019.

\bibitem{wells2022s}
B.~Wells-Edwards, ``What's in a voice? the legal implications of voice
  cloning,'' {\em Ariz. L. Rev.}, vol.~64, p.~1213, 2022.

\bibitem{coburn2022enhanced}
C.~Coburn, K.~Williams, and S.~R. Stroud, ``Enhanced realism or ai-generated
  illusion? synthetic voice in the documentary film roadrunner,'' {\em Journal
  of Media Ethics}, vol.~37, no.~4, pp.~282--284, 2022.

\bibitem{dale2022voice}
R.~Dale, ``The voice synthesis business: 2022 update,'' {\em Natural language
  engineering}, vol.~28, no.~3, pp.~401--408, 2022.

\bibitem{wang2017tacotron}
Y.~Wang, R.~Skerry-Ryan, D.~Stanton, Y.~Wu, R.~J. Weiss, N.~Jaitly, Z.~Yang,
  Y.~Xiao, Z.~Chen, S.~Bengio, {\em et~al.}, ``Tacotron: Towards end-to-end
  speech synthesis,'' {\em arXiv preprint arXiv:1703.10135}, 2017.

\bibitem{skerry2018towards}
R.~Skerry-Ryan, E.~Battenberg, Y.~Xiao, Y.~Wang, D.~Stanton, J.~Shor, R.~Weiss,
  R.~Clark, and R.~A. Saurous, ``Towards end-to-end prosody transfer for
  expressive speech synthesis with tacotron,'' in {\em international conference
  on machine learning}, pp.~4693--4702, PMLR, 2018.

\bibitem{wang2018style}
Y.~Wang, D.~Stanton, Y.~Zhang, R.-S. Ryan, E.~Battenberg, J.~Shor, Y.~Xiao,
  Y.~Jia, F.~Ren, and R.~A. Saurous, ``Style tokens: Unsupervised style
  modeling, control and transfer in end-to-end speech synthesis,'' in {\em
  International conference on machine learning}, pp.~5180--5189, PMLR, 2018.

\bibitem{jia2018transfer}
Y.~Jia, Y.~Zhang, R.~Weiss, Q.~Wang, J.~Shen, F.~Ren, P.~Nguyen, R.~Pang,
  I.~Lopez~Moreno, Y.~Wu, {\em et~al.}, ``Transfer learning from speaker
  verification to multispeaker text-to-speech synthesis,'' {\em Advances in
  neural information processing systems}, vol.~31, 2018.

\bibitem{lim2022detecting}
S.-Y. Lim, D.-K. Chae, and S.-C. Lee, ``Detecting deepfake voice using
  explainable deep learning techniques,'' {\em Applied Sciences}, vol.~12,
  no.~8, p.~3926, 2022.

\bibitem{chen2020generalization}
T.~Chen, A.~Kumar, P.~Nagarsheth, G.~Sivaraman, and E.~Khoury, ``Generalization
  of audio deepfake detection.,'' in {\em Odyssey}, pp.~132--137, 2020.

\bibitem{mcuba2023effect}
M.~Mcuba, A.~Singh, R.~A. Ikuesan, and H.~Venter, ``The effect of deep learning
  methods on deepfake audio detection for digital investigation,'' {\em
  Procedia Computer Science}, vol.~219, pp.~211--219, 2023.

\bibitem{conti2022deepfake}
E.~Conti, D.~Salvi, C.~Borrelli, B.~Hosler, P.~Bestagini, F.~Antonacci,
  A.~Sarti, M.~C. Stamm, and S.~Tubaro, ``Deepfake speech detection through
  emotion recognition: a semantic approach,'' in {\em ICASSP 2022-2022 IEEE
  International Conference on Acoustics, Speech and Signal Processing
  (ICASSP)}, pp.~8962--8966, IEEE, 2022.

\bibitem{wang2020asvspoof}
X.~Wang, J.~Yamagishi, M.~Todisco, H.~Delgado, A.~Nautsch, N.~Evans,
  M.~Sahidullah, V.~Vestman, T.~Kinnunen, K.~A. Lee, {\em et~al.}, ``Asvspoof
  2019: A large-scale public database of synthesized, converted and replayed
  speech,'' {\em Computer Speech \& Language}, vol.~64, p.~101114, 2020.

\bibitem{hennequin2020spleeter}
R.~Hennequin, A.~Khlif, F.~Voituret, and M.~Moussallam, ``Spleeter: a fast and
  efficient music source separation tool with pre-trained models,'' {\em
  Journal of Open Source Software}, vol.~5, no.~50, p.~2154, 2020.

\bibitem{mcfee2015librosa}
B.~McFee, C.~Raffel, D.~Liang, D.~P. Ellis, M.~McVicar, E.~Battenberg, and
  O.~Nieto, ``librosa: Audio and music signal analysis in python,'' in {\em
  Proceedings of the 14th python in science conference}, vol.~8, pp.~18--25,
  2015.

\bibitem{kim2021conditional}
J.~Kim, J.~Kong, and J.~Son, ``Conditional variational autoencoder with
  adversarial learning for end-to-end text-to-speech,'' in {\em International
  Conference on Machine Learning}, pp.~5530--5540, PMLR, 2021.

\bibitem{kim2018crepe}
J.~W. Kim, J.~Salamon, P.~Li, and J.~P. Bello, ``Crepe: A convolutional
  representation for pitch estimation,'' in {\em 2018 IEEE International
  Conference on Acoustics, Speech and Signal Processing (ICASSP)},
  pp.~161--165, IEEE, 2018.

\bibitem{chen2016xgboost}
T.~Chen and C.~Guestrin, ``Xgboost: A scalable tree boosting system,'' in {\em
  Proceedings of the 22nd acm sigkdd international conference on knowledge
  discovery and data mining}, pp.~785--794, 2016.

\bibitem{breiman2001random}
L.~Breiman, ``Random forests,'' {\em Machine learning}, vol.~45, pp.~5--32,
  2001.

\bibitem{fisher1936use}
R.~A. Fisher, ``The use of multiple measurements in taxonomic problems,'' {\em
  Annals of eugenics}, vol.~7, no.~2, pp.~179--188, 1936.

\bibitem{hoerl1970ridge}
A.~E. Hoerl and R.~W. Kennard, ``Ridge regression: Biased estimation for
  nonorthogonal problems,'' {\em Technometrics}, vol.~12, no.~1, pp.~55--67,
  1970.

\bibitem{bayes1763lii}
T.~Bayes, ``Lii. an essay towards solving a problem in the doctrine of chances.
  by the late rev. mr. bayes, frs communicated by mr. price, in a letter to
  john canton, amfr s,'' {\em Philosophical transactions of the Royal Society
  of London}, no.~53, pp.~370--418, 1763.

\bibitem{fix1951discriminatory}
E.~Fix and J.~Hodges, ``Discriminatory analysis, nonparametric
  discrimination,'' 1951.

\bibitem{hearst1998support}
M.~A. Hearst, S.~T. Dumais, E.~Osuna, J.~Platt, and B.~Scholkopf, ``Support
  vector machines,'' {\em IEEE Intelligent Systems and their applications},
  vol.~13, no.~4, pp.~18--28, 1998.

\bibitem{robbins1951stochastic}
H.~Robbins and S.~Monro, ``A stochastic approximation method,'' {\em The annals
  of mathematical statistics}, pp.~400--407, 1951.

\bibitem{williams1995gaussian}
C.~Williams and C.~Rasmussen, ``Gaussian processes for regression,'' {\em
  Advances in neural information processing systems}, vol.~8, 1995.

\bibitem{scikit-learn}
F.~Pedregosa, G.~Varoquaux, A.~Gramfort, V.~Michel, B.~Thirion, O.~Grisel,
  M.~Blondel, P.~Prettenhofer, R.~Weiss, V.~Dubourg, J.~Vanderplas, A.~Passos,
  D.~Cournapeau, M.~Brucher, M.~Perrot, and E.~Duchesnay, ``Scikit-learn:
  Machine learning in {P}ython,'' {\em Journal of Machine Learning Research},
  vol.~12, pp.~2825--2830, 2011.

\end{thebibliography}
\bibliographystyle{ieeetr}

\end{document}